\documentclass[showpacs,preprintnumbers,amsmath,amssymb]{revtex4}
\usepackage{graphicx}
\usepackage{dcolumn}
\makeatletter
\input{epsf}

\begin{document}
\title{High $T_c$ Superconductivity, Skyrmions and the Berry Phase}

\author{B. Basu}
\email{banasri@isical.ac.in}
\author{S. Dhar}
 \email{sarmishtha_r@isical.ac.in}
\author{P. Bandyopadhyay}
 \email{pratul@isical.ac.in}
\affiliation{Physics and Applied Mathematics Unit\\
 Indian Statistical Institute\\
 Calcutta-700108 }

\begin{abstract}
\centerline{{\bf Abstract}}

 It is here pointed out that the antiferromagnetic spin
fluctuation may be associated with a gauge field which gives rise
to the antiferromagnetic ground state chirality. This is
associated with the chiral anomaly and Berry phase when we
consider the two dimensional spin system on the surface of a $3$D
sphere with a monopole at the centre. This realizes the RVB state
where spinons and holons can be understood as chargeless spins and
spinless holes attached with magnetic flux. The attachment of the
magnetic flux of the charge carrier suggest, that this may be
viewed as a skyrmion. The interaction of a massless fermion
representing a neutral spin with a gauge field along with the
interaction of a spinless hole with the gauge field enhances the
antiferromagnetic correlation along with the pseudogap at the
underdoped region. As the doping increases the antiferromagnetic
long range order disappears for the critical doping parameter
$\delta_{sc}$. In this framework, the superconducting pairing may
be viewed as caused by skyrmion-skyrmion bound states.

\end{abstract}

\maketitle

\section{Introduction}

It is now well known that there exists an interplay between
antiferromagnetism and d-wave superconductivity in cuprate
materials. Indeed, on doping with holes, these insulating
compounds develop into superconductors even for low concentration
of holes. This implies that the antiferromagnetic spin fluctuation
plays a significant role in the development of high $T_c$
superconductivity in these materials and the d-wave
superconducting phase is a nearly antiferromagnetic Fermi liquid.
In this context Monthoux, Balatsky and Pines \cite{monp} have
considered spin fluctuation driven pairing for the cuprates near
optimal doping. Rantner and Wen \cite{ranw}, in the framework of
$U(1)$ gauge fluctuations, have studied the underdoped cuprates
where the spin behavior shows the peculiar competition between
antiferromagnetic order and singlet formation as is evidenced by
pseudogap observed in NMR and neutron scattering. The spin
pseudogap can be well explained in terms of the RVB state as
proposed by Anderson \cite{An}. It is argued  that the effect of
the preformed spin singlets present in the RVB picture on the
doped holes can be described in terms of the fact that the spin of
the doped holes becomes an excitation whereas the charge remains
tied to the empty site. This leads to the chargeless spin
excitations (spinons) and spinless charge excitations (holons).
Superconductivity arises when coherence is established after
spin-charge recombination \cite{bb1}. However, underdoped cuprates
have a peculiar property which is apparently very puzzling. As the
doping is lowered both the pseudogap and the antiferromagnetic
correlation increases. Naively, it is expected that the larger the
pseudogap stronger the spin singlet formation and weaker the
antiferromagnetic correlation. However, in the underdoped region
the scenario is different and both the pseudogap and
antiferromagnetic correlation increase.

In a study \cite{kiml} of high $T_c$ cuprates in the underdoped
region from a gauge theoretical point of view it is shown that
gauge field fluctuations effectively removes the deficiencies of
the mean field theories in explaining the antiferromagnetic
correlations as observed in experiments. It has been argued that
gauge theory with an additional coupling to holons helps to
enhance the antiferromagnetic correlations.

A model is proposed \cite{marn1,marn2} for high-$T_c$
superconductors which includes both the spin fluctuations of the
Cu$^{++}$ magnetic ions and of the spins of O$^{--}$ doped holes
(holons). The charge of the doped hole is associated to quantum
skyrmion excitations (holons) of the Cu$^{++}$ background. The
quantum skyrmion effective interaction potential is evaluated as a
function of doping and temperature indicating that Cooper pair
formation is determined by the competition between these two types
of spin fluctuations. The superconducting transition occurs when
the effective potential allows for skyrmion bound states.

In a recent paper \cite{bb1} we have also proposed a mechanism of
high $T_c$ superconductivity from the view point of chirality and
Berry phase. It is observed that the spin pairing and charge
pairing is caused by a gauge force generated by magnetic flux
quanta attached to them. Different phase structures associated
with high $T_c$ superconductivity have been studied from an
analysis of the renormalization group equation involving the Berry
phase factor $\mu$ which corresponds to the monopole strength
associated with the magnetic flux quanta. It is found that there
are two crossovers above the superconducting temperature $T_c$,
one corresponding to the glass phase and the other represents the
spin gap phase. However, the spin gap temperature $T_2^*$ is found
to be dependent on $T_c$ and $\frac{T_2^*}{T_c}$ shows a universal
behavior with respect to the hole doping $\frac{\delta}{\delta_0}$
with ${\delta_0}$ being the optimal doping rate.

In this note we shall study the topological excitations of high
$T_c$ superconductivity in cuprates in this framework and shall
show that the charge carriers appear as skyrmion excitations of
the Cu$^{++}$ spin background. The enhancement of
antiferromagnetic correlations along with pseudogap in the
underdoped region is explained. The superconducting pairing caused
by spin-charge recombination may be viewed as a consequence of
formation of skyrmion-skyrmion bound state.

In sec.2 we shall discuss spin fluctuation and RVB theory from the
view point of chirality  and Berry phase. In sec.3 we shall
discuss skyrmion excitations and the enhancement of
antiferromagnetic correlation and pseudogap in the underdoped
region. In sec.4 we shall derive the critical doping parameter
$\delta_{sc}$ for the destruction of the Neel order. In sec.5 we
shall discuss superconducting pairing in terms of skyrmions.

\section{spin fluctuation, rvb state and berry phase}
We start with a spin system which is antiferromagnetic in nature.
In terms of Schwinger bosons we may write the localized spin
$\overrightarrow{S_j}$ at site $j$ as
\begin{equation}
\overrightarrow{S_j}=\frac{1}{2}~(z^{\dag}_{j\uparrow}~,~{z^{\dag}_{j\downarrow}})\overrightarrow{\sigma}
\left(\begin{array}{c}
 z_{j\uparrow} \\
z_{j\downarrow} \end{array}\right)
\end{equation}

Here $z^{\dag}_{j\sigma}$ and $z_{j\sigma}$ represent Schwinger
bosons at site $j$ and obey boson commutative relations
$[z_{i\sigma},z^{\dag}_{j\sigma^{\prime}}]=\delta_{ij}~\delta_{\sigma
\sigma^{\prime}}$ and
$[z_{i\sigma},z_{j\sigma^{\prime}}]=[z^{\dag}_{i\sigma},z^{\dag}_{j\sigma^{\prime}}]=0$.
We have also the constraint $\sum_{\sigma} z^{\dag}_{j\sigma}~
z_{j\sigma} =1$ for $S=1/2$. The Hamiltonian for the localized
spin system is given by
\begin{equation}
H=-\frac{1}{2}~ |J|\sum_{i<j} F^{\dag}_{ij} F_{ij}
\end{equation}
where $|J|>0$ and $F_{ij}$=$\sum_{\sigma} z^{\dag}_{i\sigma}
z_{j\sigma}$.

If a hole is doped in this spin system an appreciable amount of
spin fluctuations may arise which may be represented by $Q_{ij}$
where $<Q_{ij}>=\sum_{\sigma} z_{i\sigma} z_{j\sigma}$. We may
note that the spin fluctuation $Q_{ij}$ consists of the phase
fluctuation and the amplitude fluctuation. However, as the latter
is effectively a high energy mode, so we may concentrate on the
phase fluctuation which is connected with the local gauge
transformation of $\bar{z}_{j\sigma}$ and $z_{j\sigma}$ at each
site given by
\begin{equation}
z_{j\sigma}\rightarrow z_{j\sigma}~ exp(-i \theta_j)
\end{equation}

This suggests that the transformation in the phase of $Q_{ij}$ can
be described by a gauge field, $A_{ij}$.

To visualize the spin fluctuation in a two dimensional
antiferromagnetic system we consider the Heisenberg model with
nearest neighbour interaction represented by the Hamiltonian
\begin{equation}
H = J \sum ( S^x_i S^x_j + S^y_i S^y_j + S^z_i S^z_j)
\end{equation}
where $S_i$ is a spin operator of an electron at site $i$ and
$J>0$.  The ground state of antiferromagnetic system in
$2$-dimensions on a lattice which allows frustration is
characterized by the chirality operator \cite{wieg}
\begin{equation}
W(C) = Tr \prod_{i \in C} ( \frac{1}{2} + \vec{\sigma} .
\vec{S_i})
\end{equation}
where $\sigma$ are Pauli matrices and $C$ is a lattice contour.
The topological order parameter $W(C)$ acquires the form of a
lattice Wilson loop
\begin{equation}
W(C) = e^{i \phi (c)}
\end{equation}
which may be associated with the flux represented by the gauge
field $A_{ij}$. Indeed, we may represent the chirality operator in
terms of $A_{ij}$ so that
\begin{equation} W(C)= \prod_C e^{i A_{ij}}
\end{equation}
where $A_{ij}$ represents a magnetic flux which penetrates through
a surface enclosed by the contour $C$. We may associate this
$A_{ij}$ with the phase fluctuation associated with the spin
fluctuation caused by the doped hole when we have doping induced
frustration in the system. As $A_{ij}$ represents the Berry phase
related to chiral anomaly when we describe the system in three
dimensions we may write \cite{bb1}
\begin{equation}
W(C) = e^{i 2 \pi \mu}
\end{equation}
where $\mu$ represents the monopole strength $(\hbar=c=e=1)$. In
view of this when a two dimensional frustrated spin system on a
lattice is taken to reside on the surface of a three dimensional
sphere of a large radius in a radial magnetic field, we can
associate the chirality with the Berry phase. Eventually this will
give rise to RVB state \cite{bbdp}.

It may be remarked here that when a chiral current interacts with
a gauge field, we have the anomaly which is related to the Berry
phase through the relation \cite{dbpb}
\begin{equation}
\label{chi} q~=2 \mu =- \frac{ 1}{2} \int \partial_\mu J^5_\mu
d^4x =~\frac{1}{16 \pi^2}~ Tr~ \int~^*F_{\mu\nu}F_{\mu\nu}~d^4 x
\end{equation}
where $J^5_\mu$ is the axial vector current $\bar{\psi} \gamma_\mu
\gamma_5 \psi$, $F_{\mu\nu}$ is the field strength and
$^*F_{\mu\nu}$ is the Hodge dual. Evidently $q=2\mu$ represents
the Pontryagin index.

To study the spin system leading to a RVB state we consider a
generalized nearly antiferromagnetic spin model with nearest
neighbor interaction as
\begin{equation}
  H = J \sum ( S^x_i S^x_j + S^y_i S^y_j
+ \Delta S^z_i S^z_j)
\end{equation}
where $J > 0$ and the anisotropy parameter $\Delta = \frac{2\mu +
1}{2}$ \cite{pb}.  The Berry phase factor $\mu$ can take the
values $\mu = 0, \pm 1/2, \pm 1, \pm 3/2 ........$. It is noted
that $\Delta = 1$ corresponds to $\mu = 1/2$ and represents the
isotropic Hamiltonian which is $SU(2)$ invariant. For $\Delta
\rightarrow \infty$, it corresponds to an Ising system. When
$\Delta = 0 (\mu = - 1/2)$ we have the $XX$ model. For a
frustrated spin system, this corresponds to the singlets of spin
pairs which eventually represents the RVB state giving rise to a
non-degenerate quantum liquid.

In a recent paper \cite{bb1} we have studied the different phases
associated with superconductivity in cuprates through the
renormalization group analysis involving the factor $\mu$. It is
noted that $\mu$ takes the usual discrete values of $0, \pm
\frac{1}{2}, \pm 1, \pm \frac{3}{2} ...$ at fixed points of the RG
flows where $\mu$ is stationary and represents the Berry phase
factor $\mu^\ast$ of the theory. In terms of energy scale, it is
found that as energy increases (decreases) $\mu$ also increases
(decreases). So to study a critical phenomena, we can associate a
critical temperature with a standard discrete value of $\mu$
corresponding to the Berry phase factor $\mu^\ast$ which
represents a fixed point of the RG flows. To study the crossover,
it is  noted that for $0 \le |\Delta| < 1$ there are three
critical values corresponding to $\mu = 0$, $\mu = - \frac{1}{2}$
and $\mu=-1$
 which
represent the fixed points of the RG flow. We associate three
critical temperatures $T^*_1$, $T^*_2$ and $T_c$ with fixed values
of $\mu=0$, $\mu=- \frac{1}{2}$ and $\mu=-1$ respectively.
However, in a frustrated spin system, the chirality demands that
$\mu$ should be non-zero. So the critical value $\mu = 0$ is not
achieved and as such there will be random coupling around the
value $\mu = 0$. This will then represent the cluster glass phase
at this critical temperature $T_1^*$.
 In this situation, after doping, holes will form a glass of
 stripes.
The next crossover will be at $ \mu=- \frac{1}{2}$ corresponding
to the pseudogap (spin gap) phase. As $\mu=-\frac{1}{2}$
corresponds to  $\Delta=0$, the  spin chain will represent the
system of spin singlets leading to RVB phase. The spin-charge
separation here describes the spin gap (pseudogap) phase. Finally,
we arrive at the superconducting transition temperature $T_c$ at
$\mu=-1$ corresponding to $\Delta=-1/2$. At this point, the Ising
part coupling constant is $\frac{-1}{2}J$ with a sign change which
represents an attractive force causing the superconducting pair
formation.

The concentration of doped holes may be parameterized by a length
scale $L$. In view of this, we may consider $\mu$ as a function of
$\delta$ at a fixed temperature. The doped holes will suppress the
$U(1)$ gauge fluctuation describing the antiferromagnetic spin
fluctuation. At zero doping, we have the Heisenberg
antiferromagnet. The Neel temperature $T_N$ is reduced upon doping
and at a critical doping $T_N(\delta_c)=0$. As the doping is
increased, the magnetic long range order is destroyed. However, as
the doping is lowered both the pseudogap and the antiferromagnetic
correlation is increased. This aspect will be discussed in the
next section.

\section{Skyrmions, Antiferromagnetic correlation and pseudogap}
To study the spinon and holon excitations in our model
\cite{bb1,bb2,bb3} let us consider a single spin down electron at
a site $j$ surrounded by an otherwise featureless spin liquid
representing a RVB state. Due to the chirality caused by the gauge
fluctuation we may consider the system such that a monopole
represented by $\mu=-1/2$ is in the background leading to RVB
ground state. As a result, the single spin will be characterized
by $|\mu|=1$ formed by the single spin state characterized by
$\mu=-1/2$ coupled with the orbital spin $\mu=-1/2$ caused by the
monopole in the background. This neutral spin  attached with
magnetic flux quanta given by $|\mu|=1$ will appear as an
excitation and represent the spinon. Now when a doped hole
interacts  with this spinon, it will give rise to a spinless
charged excitation called holon. Thus holons may also be
represented by $|\mu|=1$ characterized by a flux
$\phi_0=\frac{hc}{2e}$. The residual spinon will then correspond
to $\mu_{eff}=0$ which is realized when the unit of magnetic flux
characterized by $\mu=-1/2$ associated with the single down spin
in the RVB liquid forms a pair with another up spin having
$\mu=+1/2$ associated with the hole. Again the holon having
$|\mu_{eff}|=1$ will also eventually form a pair each
characterized by $|\mu|=1/2$. Indeed for any integer $\mu$ the
Berry phase may be removed to the dynamical phase and the
geometric phase is realized when a pair is formed \cite{bbp}. Thus
the spinon and holon may be viewed as if a neutral spin as well as
a charged spinless hole is attached with a magnetic flux quantum
characterized by $|\mu|=1/2$ and these appear in a pair.

Now it is noted that  when a spinless hole is dressed with a
magnetic flux quantum given by $|\mu|=1/2$, this will represent a
skyrmion. Indeed, the magnetic flux quantum has its origin in the
background chirality which is associated with the chiral anomaly
and Berry phase. Indeed, from  eqn.(\ref{chi}), we note that the
Berry phase factor $\mu$ is associated with
$^*F_{\mu\nu}F_{\mu\nu}$ and we can write
\begin{eqnarray}
\label{pont}
 q~=&&~2\mu\nonumber\\
 =&&-{\frac{1}{16 \pi^2}}~\int Tr ~^{*} {{F}}_{\mu\nu} {{F}}_{\mu\nu} d^4 x\nonumber\\
=&&\int d^4 x ~\partial_\mu \Omega_\mu
\end{eqnarray}
where
\begin{equation}
\Omega^{\mu} = -{\frac{1}{16 \pi^2}} \epsilon^{\mu \nu \alpha
\beta}
 ~Tr(A_{\nu} F_{\alpha \beta} + {\frac{2}{3}} A_{\nu}A_{\alpha}A_{\beta})
\end{equation}
is the Chern-Simons secondary characteristic class. In case we
have $F_{\alpha\beta}=0$ we can write
\begin{equation}
A_\mu=g^{-1} \partial_\mu g,~~~~~~~~~~~     g \in SU(2)
\end{equation}
and $\Omega_\mu$ will represent a topological current $J_\mu$
given by
\begin{equation}
J_\mu~=~ {\frac{1}{24 \pi^2}}~ \epsilon^{\mu\nu\alpha\beta}
~Tr(g^{-1}\partial_{\nu}g)(g^{-1}\partial_{\alpha}g)
(g^{-1}\partial_{\beta}g)
\end{equation}

This may be written in terms of chiral fields $\pi_a~(a=0,1,2,3)$.
\begin{equation} J_\mu~=~{\frac{1}{12 \pi^2}}
\epsilon^{\mu\nu\alpha\beta} \epsilon^{abcd} \pi_a
\partial_{\nu} \pi_b \partial_{\alpha} \pi_c \partial_{\beta} \pi_d
\end{equation}

Now representing a hole by a Dirac fermion field $\psi$ we may
consider the doped hole coupling with the magnetic flux associated
with the chirality in terms of the interaction given by the
Lagrangian
\begin{equation}
L ={\bar \psi}(i{\hat D}+ im(\pi_0 +i\gamma_5 {\vec \pi}{\vec
\tau}))\psi
\end{equation}
where ${\hat D}$=$ \gamma_\mu(\partial_\mu - iA_\mu)$ following
the constraint $\pi^2_0+{\vec \pi}^2=1$

The Dirac fermion may be viewed as if it has flavor $N$ so that
for polarized and unpolarized state we have $N=1$ and $2$
respectively. Now integrating for fermions, we can write the
action
\begin{eqnarray}
\label{act}
W~=&&~-~ln~\int exp(-L d^4 x) D\psi~D\bar{\psi}\nonumber\\
=&&~-~N~ln~ {\bf{Det}}(i\hat{D}+im g^{\gamma_5})\nonumber\\
=&&~i~N~\int d^4 x A_\mu J_\mu ~ + i\pi N H_3\nonumber\\
&+&NM^2 \int d^4 x ~Tr~ (\partial_\mu g^{-1}\partial_\mu g)
\end{eqnarray}
Here $g^{\gamma_5}=\frac{1+\gamma_5}{2} g +\frac{1-\gamma_5}{2}
g^{-1}$. $M$ is a coupling constant having dimension of mass.
$H_3$ is a topological invariant of the map of the space-time into
the target space $S^3$. There are only two homotopy  classes
$\pi_4(S^3)=Z_2$, so that $H_3=0$ or $1$. In fact the term $i \pi
H_3$ is the geometric phase and represents the $\theta$-term. Thus
we see that the charge carriers dressed with magnetic flux can be
represented by a nonlinear $\sigma$-model and may be treated as
skyrmions.

To study the underdoped region of cuprates in this framework, we
note that spinon-holon interaction through the gauge force
effectively leads to a spin pair characterized by $\mu_{eff}=0$
where the isolated down spin in the background with $\mu=-1/2$
forms the pair with the up spin of the hole with $\mu=+1/2$.
Indeed this may be taken to represent as a spinon-antispinon bound
state. This essentially corresponds to the SF flux phase as
suggested by Rantner and Wen \cite{ranw}. Indeed we can visualize
a spin as a massless fermion and this picture of spinon-holon
interaction may correspond to a massless fermion coupled to $U(1)$
gauge field along with the holons coupled with the gauge field.
The pair formed by massless fermions (spins) dressed with magnetic
flux may be viewed as a spinon-antispinon bound state. This
spinon-antispinon bound state present in the nearly
antiferromagnetic chain will enhance the antiferromagnetic
correlation of the system. The simultaneous presence of spin
singlet state will lead to the pseudogap (spin gap). Thus in the
underdoped region we will have the enhancement of the
antiferromagnetic correlation along with the pseudogap. As
mentioned earlier, as doping increases, the antiferromagnetic long
range order is destroyed.

\section{skyrmions, critical doping and the destruction of the antiferromagnetic
order} In the present framework, superconductivity arises with the
charge spin recombination when a phase coherence is established.
Indeed, prior to spin-charge recombination, a spinless holon may
be viewed as if a spinless hole is moving in the background of a
monopole. This eventually causes the hole pair formation each
having a magnetic flux quantum characterized by $|\mu|=1/2$. When
the spin charge recombination occurs a spin pair each having unit
magnetic flux interact with each other through a gauge force and a
phase coherence is established. As we have pointed out in the
earlier section that the charge carrier attached with a magnetic
flux corresponds to a skyrmion, we may view the superconducting
pair as
 a skyrmion-skyrmion bound state. Indeed, the
skyrmion excitation is created at each position of the carriers
and plays a role of magnetic field for the carriers. Because of
the magnetic field around a carrier, the Lorentz force acts on
another carrier. Due to this Lorentz force an attractive
interaction is induced between carriers and leads to Cooper pair
formation.

It is noted that the mechanism  suggests a d-wave pairing. As
already pointed out by Kotliar and Liu \cite{kotl} that in the RVB
theory  spinons form the d-wave pairing. Now in the
superconducting pair, the spin charge recombination occurring
through spinon-holon interaction along with the phase coherence
suggests the charge carriers also have d-wave pairing. Indeed, the
fact that superconductivity occurs in the vicinity of
antiferromagnetic long range order, the Cooper pair is d-wave.

It is known that skyrmion topological defects which are introduced
by doping are responsible for the destruction of the
antiferromagnetic order and their energy may be used as an order
parameter \cite{marn1,marn2}. Indeed, in two spatial dimensions
the nonlinear sigma field $n^a$ may be expressed in the $CP^1$
language in terms of a doublet of complex scalar fields $z_i,~
i=1,2$ with the component $z^{\dag}_i z_i=1$ as
\begin{equation}
n^a=z^{\dag}_i \sigma^a_{ij} z_j
\end{equation}
where $\sigma^a$ are Pauli matrices. In this language the continuous field theory
corresponding to the Heisenberg antiferromagnet is described by the Lagrangian
density in $2+1$ dimensions
\begin{equation}
\label{lns} L_{ns}=(D_\mu z_i)^{\dag} (D_\mu z_i)
\end{equation}
where $D_\mu=\partial_\mu + iA_\mu$ and ${\cal A}_\mu=iz^{\dag}_i
\partial_\mu z_i$. This possesses solitonic solutions
called skyrmions and charge is defined as
\begin{equation}
Q=\int d^3 x J^0
\end{equation}
where $J^0$ is the zero-th component of the topological current $J^\mu=\frac{1}{2\pi}\epsilon^{\mu\alpha\beta}\partial_\alpha {\cal A}_\beta$.
It is noted that $Q$ is nothing but the magnetic flux of the field ${\cal A}_\mu$ indicating
that skyrmions are vortices and represent defects in the ordered Neel state.

Now the following Lagrangian density may be proposed for describing the dopants and their interaction
with the background lattice in $2+1$ dimensions with the topological $\theta$-term
\begin{equation}
\label{lag} L_{z,\psi}=(D_\mu z_i)^{\dag} (D^\mu
z_i)+i\bar{\psi_a}\partial_\mu \gamma_\mu \psi_a -m^* v_F
\bar{\psi_a} \psi_a -\bar{\psi_a}\partial^\mu \psi_a {\cal A}_\mu
+ L_H
\end{equation}
where the hole dopants are represented by a two-component Dirac
field $\psi_a,~ m^*$ and $v_F$ are respectively the effective mass
and Fermi velocity of dopants. Here $L_H$ is the Hopf term given
by
\begin{equation}\label{lagH}
  L_H=\frac{\theta}{2}\epsilon^{\mu\alpha\beta} {\cal A}_\mu \partial_\alpha {\cal A}_\beta
\end{equation}
It should be mentioned here that long ago it was shown that
{\cite{frad} antiferromagnetic spin correlation do not produce a
Hopf term on a two dimensional square lattice. In fact, these
authors have pointed out that the presence of the nontrivial Hopf
term may come from something else other than the spin themselves.
In this case, the Hopf term arises from the doped holes which will
be revealed later.

 It is noted that the dopant dispersion relation is given by
\begin{equation}
  \epsilon (k)=\sqrt{k^2 v_F^2 + (m^* v_F^2)^2}
\end{equation}
which is valid for $YBCO~(YBa_2 Cu_3O_{6+\delta})$ where the Fermi
surface has an almost circular shape which is centered at ${\bf k
}=0$. For $LSCO$ $(La_{2-\delta} Sr_\delta CuO_4)$ the Fermi
surface is different \cite{marn2} which corresponds to a
dispersion relation of the form
\begin{equation}
\epsilon (k)= \sqrt{[(k_x \pm \frac{\pi}{2})^2 + (k_y \pm
\frac{\pi}{2})^2] v_F^2+(m^* v_F^2)^2}
\end{equation}

Now following Marino \cite{marn2} the doping parameter $\delta$ is
introduced by means of a constraint in the fermion integration
measure
\begin{equation}
D[\bar{\psi_a},\psi_a]= D\bar{\psi_a} D \psi_a ~\delta
(\bar{\psi_a}\gamma_\mu \psi_a -\Delta^\mu)
\end{equation}
where $\Delta^\mu=4 \delta \int^{\infty}_{x,L} d \xi^\mu~
\delta^3(z-\xi)$ for a dopant at the position $x$ and varying
along the line $L$. Here the factor $4$ corresponds to the
degeneracy of the representation ($4$-component) for the Fermi
fields. This yields the partition function
\begin{eqnarray}
\label{part}
Z=&&~\int D (\bar{z_0},z,{\cal A},\bar{\psi},\psi)~\delta(\bar{z}z-1)~\delta(\bar{\psi}\gamma_\mu \psi-\Delta^\mu)\nonumber\\
&&~\times exp~\{\int^{\infty}_0 d^3 x [~2 \rho_s (D_\mu z^{\dag}_i
D_\mu z_i)+\bar{\psi}(i \partial_\mu \gamma_\mu -\frac{m^* v
_F}{\hbar}-\gamma^\mu {\cal A}_\mu) \psi + L_H]\}
\end{eqnarray}
where $\rho_s$ is the spin stiffness and $L_H$ is the Hopf term.

Upon integration over the fields $z,~\bar{z},~\bar{\psi},~\psi$
the resulting equation of motion for the zero-th component ${\cal
A}_0$ yields the result
\begin{equation}
  \theta~\epsilon^{ij} \partial_i A_j=4\delta~\delta^2~({\bf
  z}-{\bf x}(t))
\end{equation}
where ${\bf x}(t)$ is the dopant position at a time $t$. If $B$ is
the {\it magnetic flux} or vorticity of ${\cal A}_\mu$ then this
equation becomes
\begin{equation}
\label{theta} \theta B=4\delta~\delta^2~({\bf z}-{\bf x}(t))
\end{equation}

For the skyrmion $B=\delta^2~({\bf z}-{\bf x}(t))$ indicates that
the skyrmion topological defect configuration coincides with the
dopant position at any time and \cite{marn1}
\begin{equation}\label{mari}
\pi \theta~=~2\delta
\end{equation}

When we translate this result in the $3+1$ dimensional formalism
where the $2$D spin system is considered to reside on the surface
of a $3$D sphere with a monopole at the centre, we note that in
the Lagrangian (\ref{lag}), apart from $\mu$ being a $4$
dimensional index, we have to replace the Hopf term by the
topological Pontryagin term given by
\begin{equation}
  P=-{\frac{1}{16 \pi^2}}~ ^{*} {\mathcal{F}}_{\mu\nu}
{\mathcal{F}}_{\mu\nu}
\end{equation}
where
\begin{equation}
\label{star} ^{*}{\mathcal{F}}_{\mu\nu}
=\frac{1}{2}~\epsilon^{\mu\nu\alpha\beta}{\cal F}_{\alpha\beta}
\end{equation}

It is noted that in the partition function (\ref{part}) when $\int
L_H d^3 x$ is replaced by $\int P d^4 x$, the latter integral just
represents the Pontryagin index $q$ related to the monopole
strength $\mu$ through the relation $q=2\mu$ as given by
eqn.(\ref{pont}).

From dimensional hierarchy, the relation between topological terms
suggests that in $3+1$ dimensions, when $L_H$ is replaced by
$L_P$, the coefficient $\theta$ is related to $\mu$. Indeed
replacing $L_H$ by the Chern-Simons Lagrangian
\begin{equation}\label{lagcs}
L_{cs}=\frac{k}{4\pi}\epsilon^{\mu\alpha\beta} {\cal A}_\mu
\partial_\alpha {\cal A}_\beta
\end{equation}
we note that the current is given by
\begin{equation}
\label{cur} J_\mu =\frac{k}{2 \pi}
\epsilon^{\mu\alpha\beta}\partial_\alpha {\cal A}_\beta
\end{equation}
and the zeroth component corresponds to
\begin{equation}
\label{curr} J_0=k\frac{B}{2 \pi}
\end{equation}

So from the relation(\ref{lagH}), (\ref{lagcs}) and (\ref{mari})
we find
\begin{equation}
\pi \theta=\frac{k}{2}=2\delta
\end{equation}

It is noted that if we take $\delta=0$ which represents the pure
undoped quantum antiferromagnet we do not have the Hopf term which
is consistent with the observation of Fradkin and Stone
\cite{frad}

It has been shown in ref.\cite{pb} that the Chern-Simons
coefficient $k$ is related to the monopole strength $\mu$ in $3+1$
dimensions by the relation $k=2\mu$. This implies $\mu=2\delta$.
 As in the previous section we have noted
that each charge carrier in the superconducting pair is associated
with the skyrmion topological defect which is caused by the
magnetic flux quantum having $|\mu|=1/2$, superconductivity occurs
at $T=0$ for the critical doping parameter $\delta_{sc}$ given by
$|\mu|=1/2=2\delta_{sc}$ yielding $\delta_{sc}=.25$ for $YBCO$.
When the doping parameter $\delta$ is connected with the oxygen
stoichiometry parameter $x$ we have the  relation $\delta=x-.18$
so that we have $x_{sc}=.43$, which is in good agreement with the
experimental value $x_{sc}=.41 \pm .02$ \cite{exp1,marn2}. For
$LSCO$, the Fermi surface has four branches and this yields
$\delta_{sc}=x_{sc}=.06$ which is to be compared with the
experimental result $x_{sc}=.02$ \cite{exp2}. It is noted that
$\delta_{sc}$ is a universal constant depending only on the nature
of the Fermi surface.

We have pointed earlier out that in $3+1$ dimensions chiral
anomaly leads to the realization of fermions represented by doped
holes interacting with chiral boson fields $\pi_i$, with the
constraint $\pi^2_0+{\vec \pi}^2=1$. The mapping of the space-time
manifold on the target space leads to the homotopy
$\pi_4(S^3)=Z_2$ which takes the values $0$ or $1$ and leads to
the $\theta$-term representing the geometric phase. The third term
in eqn.(\ref{act}) gives rise to the solitonic solution such that
the charge carrier appears as a skyrmion. However in $3+1$
dimensions, the stability of the soliton is not generated by this
term alone as rescaling of the scale variable $x\rightarrow\lambda
x$ may lead to shrinking it to zero size. However, in the present
framework, the attachment of magnetic field with the charge
carrier will prevent it from shrinking it to zero size.

Indeed this gives rise to a gauge theoretic extension of the
extended body so that the position variable may be written as
\begin{equation}
Q_\mu ~=~q_\mu~+~iA_\mu
\end{equation}
where $q_\mu$ is the mean position. As $\mu=-1/2$ and $+1/2$
corresponds to vortices in the opposite direction we may consider
$A_\mu$ as $SU(2)$ gauge field when the field strength is given by
\begin{equation} F_{\mu\nu} = \partial_{\mu} A_{\nu} -
\partial_{\nu} A_{\mu} + [A_{\mu}, A_{\nu}] \end{equation}
where $A_{\mu}$ is a $SU(2)$ gauge field. When $F_{\mu\nu}$ is
taken to be vanishing at all points on the boundary $S^3$ of a
certain volume $V^4$ inside which $F_{\mu\nu} \neq 0$, in the
limiting case towards the boundary, we can take
\begin{equation} A_{\mu} = g^{-1}
\partial_{\mu} g, ~~~~~~ g \in SU(2) \end{equation}

This helps us to write the action incorporating the $\theta$ -term
as
\begin{equation}
\label{action}
\begin{array}{lcl}
S &=&\displaystyle{\frac{M^2}{16} \int Tr(\partial_{\mu} g^{-1}
\partial_{\mu} g) d^4 x + \frac{1}{32 \eta^2} \int Tr [\partial_{\mu} g
g^{-1}, \partial_{\nu} g g^{-1}]^2 d^4 x}\\
&&\displaystyle{+ {\frac{i\pi}{24 \pi^2}}\int_{S^3} dS_{\mu}
\epsilon^{\mu\nu\lambda\sigma}
Tr[(g^{-1} \partial_{\nu} g)(g^{-1} \partial_{\lambda} g)(g^{-1} \partial_{\sigma} g)]}\\
\end{array}
\end{equation}
where $M$ is a constant having the dimension of mass and $\eta$ is
a dimensionless coupling constant. Here the first term is related
to the gauge noninvariant term $M^2 A_{\mu} A^{\mu}$, the second
term (Skyrme term) is the stability term which arises from the
term $F_{\mu\nu}F^{\mu\nu}$ and the third term is the $\theta$
-term given by $^{*}F_{\mu\nu}F_{\mu\nu}$ which is related to the
chiral anomaly and Berry phase.

Marino and Neto \cite{marn2} have pointed out that at the critical
doping $\delta_{sc}$, the energy of the skyrmion vanishes. When we
compute the energy of the skyrmion from the action (\ref{action}),
we find the expression for the minimum energy \cite{skyr} as

\begin{equation}
E_{min}=\frac{12 \pi^2 M}{\eta}
\end{equation}
and the size for $E_{min}$ as
\begin{equation}
R_0=\frac{1}{2M \eta}
\end{equation}
Taking $M$ and $\eta$ as a function of $\delta$, we note that for
the vanishing energy we have $M(\delta_{sc})=0$ which corresponds
to the fact that the spin stiffness vanishes. From the relation
for $R_0$, it indicates that the skyrmion size is infinite.
However, we can have the vanishing energy for finite nonzero
$M(\delta)$ when $\eta$ is infinite. This suggests that at this
point $R_0=0$. This implies that for finite $M$, the vanishing
energy suggests that the skyrmion shrinks to the zero size. So
apart from energy, we can take the size of the skyrmion also as an
order parameter.

\section{discussion}
It has been pointed out here that the antiferromagnetic spin
fluctuation gives rise to a gauge field which determines the
antiferromagnetic ground state chirality. This is related to the
Berry phase and helps us to realize the RVB state where spinons
and holons can be understood as chargeless spins and spinless
holes attached with magnetic flux. The attachment of the magnetic
flux of the charge carrier suggests that this may be viewed as a
skyrmion. The interaction of a massless fermion representing a
neutral spin with a gauge field along with the interaction of a
spinless hole with the gauge field enhances the antiferromagnetic
correlation along with the pseudogap at the underdoped region. The
superconducting pairing may be viewed as caused by
skyrmion-skyrmion bound states. This effectively leads to
topological superconductivity. It is also shown that the
destruction of antiferromagnetic order is at the critical doping
parameter $\delta_{sc}$ which is a universal constant depending on
the nature of the Fermi surface.

Abanov and Wiegman \cite{abaw,abaw1} have pointed out that
topological superconductivity in $3+1$ dimensions and $2+1$
dimensions has its roots in the $1$D Peierls-Fr\"{o}hlich model
which suggests that the $2\pi$ phase solitons of the Fr\"{o}hlich
model \cite{fro} are charged and move freely through the system
making it an ideal conductor. In spatial dimension greater than
one this corresponds to superconductivity when the solitonic
feature of a charge carrier is attributed to the attachment of a
magnetic flux to it. It may be remarked here that in $1+1$
dimensions we will have a nonlinear sigma model with the
Wess-Zumino term when the target space is $S^3$ which is the
$O(4)$ nonlinear sigma model. In the Euclidean framework however,
this geometrically corresponds to the attachment  of a vortex line
to the two dimensional sheet which is topologically equivalent to
the attachment of a magnetic flux \cite{royb}. This suggests that
the topological feature of ideal conductivity visualized by
Fr\"{o}hlich in $1+1$ dimensions and that of superconductivity in
$2+1$ and $3+1$ dimensions have a common origin.


\newpage

\begin{thebibliography}{*}
\bibitem{monp} P. Monthoux, A. Balatsky and D. Pines, Phys. Rev. Lett. {\bf 67}, 3448
(1991).\\
see also:\\
A. V. Chubukov and J. Schmalian, Phys. Rev. {\bf B 57}, R11085
(1998).\\
J. Schmalian, D. Pines and B. P. Stojkovi\'{c}, Phys. Rev.
Lett. {\bf 80}, 3839 (1998).\\
T. Moriya, K. Ueda and Y. Takahashi, J. Proc. Soc. Jpn. {\bf 49},
2905 (1990).\\
T. S. Nunner, J. Schmalian and K. H. Bennemann, Phys. Rev. {\bf B
59}, 8859 (1999).
\bibitem{ranw} W. Rantner and X. G. Wen, Phys.
Rev. {\bf B 66}, 144501 (2002).
\bibitem{An} P. W. Anderson, Science {\bf 235}, 1196 (1987).
\bibitem{bb1} B. Basu, P. Bandyopadhyay and D. Pal, Int. J. Mod. Phys. {\bf B
17}, 293 (2003).
\bibitem{kiml} D. H. Kim and P. A. Lee, Annals Phys. {\bf 272}, 130
(1999).
\bibitem{marn1} E. C. Marino, Phys. Lett. {\bf A 263}, 446 (1999).
\bibitem{marn2} E. C. Marino and M. B. Neto, Phys. Rev. {\bf B 64},
092511 (2001).

\bibitem{wieg} P. Wiegmann, Prog. Theor. Phys. Suppl. {\bf
101}, 243 (1992);  {\it Topological Mechanism \\
of
Superconductivity in Field Theory, Topology and Condensed Matter
Physics,}\\ ed. H. B. Geyer, Springer, Berlin, (1999).
\bibitem{bbdp} B. Basu, D. Pal and P. Bandyopadhyay, Int. J. Mod. Phys. {\bf B
13}, 3393 (1999).
\bibitem{dbpb} D. Banerjee and P. Bandyopadhyay, J. Math. Phys. {\bf 33}, 990
(1992).

\bibitem{pb} P. Bandyopadhyay, Int. J. Mod. Phys. {\bf A 15}, 1415
(2000).
\bibitem{bb2} D. Pal, B. Basu and P. Bandyopadhyay, Phys.
Lett. {\bf A 299}, 304 (2002).
\bibitem{bb3} D. Pal and B. Basu, Europhys. Lett. {\bf 56}(1), 99
(2001).
\bibitem{bbp} B. Basu, D. Banerjee and P. Bandyapadhyay,  Phys. Lett.
{\bf A 236}, 125 (1997).
\bibitem{kotl} G. Kotliar and J. Liu, Phys. Rev. {\bf B 38}, 5142
(1988).
\bibitem{frad} E. Fradkin and M. Stone, Phys. Rev. {\bf B 38},
7215 (1988).
\bibitem{exp1} J. Rossat-Mignod {\it et. al., Dynamics of Magnetic
Fluctuations in High-Temperature\\
 Superconductors}, G. Reiter, P.
Horsch and G. C. Psaltakis, Eds., Plenum, NY, (1991).
\bibitem{exp2}F. Borsa {\it et. al.}, Phys. Rev. {\bf B 52},
7334 (1995).
\bibitem{skyr} B. Basu, S. Dhar and P. Bandyopadhyay, preprint
cond-mat/0208426.
\bibitem{abaw} A. G. Abanov and P. B. Wiegmann, Phys. Rev. Lett. {\bf 86}, 1319
(2001).
\bibitem{abaw1} A. G. Abanov and P. B. Wiegmann, Nucl. Phys. {\bf B 570}, 685 (2000).
\bibitem{fro} G. Fr\"{o}hlich, Proc. Roy. Soc. {\bf A 223}, 296 (1954).

\bibitem{royb} A. Roy and P. Bandyopadhyay, J. Math. Phys. {\bf 33}, 1178 (1992).
\end{thebibliography}
\end{document}